\documentclass[twocolumn,nofootinbib,amsmath,amssymb,aps,prl,balancelastpage,superscriptaddress, floatfix]{revtex4-1}

\usepackage{amsmath}
\usepackage{amsfonts}
\usepackage{bm}
\usepackage{graphicx}
\usepackage{hyperref}
\usepackage{soul}
\usepackage{color}
\usepackage{comment}

\enlargethispage{\baselineskip}

\hypersetup{colorlinks   = true,
            urlcolor     = blue,
            citecolor    = blue,
            linkcolor    = blue,
            menucolor    = blue,
            anchorcolor  = blue,
            filecolor    = blue}
\everymath{\displaystyle}

\begin{document}

\title{Have pulsar timing array methods detected a cosmological phase transition?}

\author{Andrea Addazi}
\email{addazi@scu.edu.cn}
\affiliation{Center for Theoretical Physics, College of Physics Science and Technology, Sichuan University, 610065 Chengdu, China}
\affiliation{INFN, Laboratori Nazionali di Frascati, Via E. Fermi 54, I-00044 Roma, Italy, EU}

\author{Yi-Fu Cai}
\email{yifucai@ustc.edu.cn}
\affiliation{Deep Space Exploration Laboratory/School of Physical Sciences,
University of Science and Technology of China, Hefei, Anhui 230026, China}
\affiliation{CAS Key Laboratory for Researches in Galaxies and Cosmology/Department of Astronomy,
School of Astronomy and Space Science, University of Science and Technology of China, Hefei, Anhui 230026, China}

\author{Antonino Marcian\`o}
\email{marciano@fudan.edu.cn}
\affiliation{Center for Field Theory and Particle Physics \& Department of Physics, Fudan University, 200433 Shanghai, China}
\affiliation{INFN, Laboratori Nazionali di Frascati, Via E. Fermi 54, I-00044 Roma, Italy, EU}

\author{Luca Visinelli}
\email{luca.visinelli@sjtu.edu.cn\\ }
\affiliation{Tsung-Dao Lee Institute (TDLI), 520 Shengrong Road, 201210 Shanghai, China} 
\affiliation{School of Physics and Astronomy, Shanghai Jiao Tong University, 800 Dongchuan Road, 200240 Shanghai, China}

\begin{abstract}
\noindent
We show that the recent detection of a gravitational wave (GW) background reported by various pulsar timing array (PTA) collaborations including NANOGrav-15yr, PPTA, EPTA, and CPTA can be explained in terms of first order phase transitions (FOPTs) from dark sector models (DSM). Specifically, we explore a model for first order phase transitions that involves the majoron, a Nambu-Goldstone boson that is emerging from the spontaneous symmetry breaking of a $U(1)_{L}$ or $U(1)_{B-L}$ symmetry. We show how the predicted GW power spectrum, with a realistic choice of the FOPT parameters, is consistent with 1-$\sigma$ deviations from the estimated parameters of the background detected by the PTA collaborations.
\end{abstract}
\maketitle

\textbf{Introduction.}
Fresh batches of pulsar timing (PTA) data have recently been released by various collaborations, reporting an excess of a stochastic spectrum which appears to be compatible with a gravitational wave background (GWB) at frequencies around $f\sim 1\div 10\, {\rm nHz}$. Among these, the North American Nanohertz Observatory for Gravitational Waves (NANOGrav) collaboration has recently released the data collected in the first $15{\rm\,yrs}$ of activity~\cite{NANOGrav:2023gor, NANOGrav:2023icp, NANOGrav:2023hfp, NANOGrav:2023ctt, NANOGrav:2023hvm}. In addition to previous analyses~\cite{NANOGrav:2020bcs, Chen:2021rqp, Goncharov:2021oub, Antoniadis:2022pcn}, the recent NANOGrav 15-year data also reports the evidence for quadrupolar correlations that follow the signature described by Hellings and Downs~\cite{Hellings:1983fr} and is then conclusive with regard to the nature of the detection. Similar evidences have been independently reported by the European Pulsar Timing Array (EPTA)~\cite{Antoniadis:2023lym, Antoniadis:2023puu, Antoniadis:2023ott, Antoniadis:2023aac, Antoniadis:2023xlr, Smarra:2023ljf}, the Parkes Pulsar Timing Array (PPTA)~\cite{Zic:2023gta, Reardon:2023zen, Reardon:2023gzh}, and the Chinese Pulsar Timing Array (CPTA)~\cite{Xu:2023wog}, pointing at a broadly consistent picture.

An intriguing possibility consists in the gravitational waves background (GWB) to be released primordially during a first order phase transition (FOPT). This scenario, taking place in the early Universe, was suggested by three of us~\cite{Addazi:2017nmg,Addazi:2020zcj} and later in Ref.~\cite{Nakai:2020oit}. We show that our predictions of the FOPT parameters are consistent with 1-$\sigma$ deviations from the parameters estimated by the GWB detected by NANOGrav, so that the FOPTs arising from a dark sector provides a natural explanation to the NANOGrav detection without being in conflict with any other phenomenological bounds. We explore in detail a FOPTs model that involves the majoron, a Nambu-Goldstone boson that is emerging from the spontaneous symmetry breaking of a $U(1)_{L}$ or $U(1)_{B-L}$ symmetry~\cite{Addazi:2017oge,Addazi:2017nmg}. 

This interpretation of the experimental data clearly requires the gravitational wave spectral information to be compatible with the GWB assumption. Reliable models on other possible astrophysical sources, including extragalactic ones, enable to distinguish the origin of the GWB --- see e.g.\ Refs.~\cite{Schneider:2010ks, Kuroyanagi:2018csn}. Future observations of the GWB will enables to probe astrophysical models on their origin, exploring frequency regions even below the nHz regime, motivated by known physics~\cite{Neronov:2020qrl, Moore:2021ibq, Brandenburg:2021tmp, DeRocco:2023qae} and other exotic models~\cite{Chang:2019mza, Ramberg:2019dgi, Freese:2023fcr}.

\textbf{FOPTs phenomenology of the majoron.} 
We consider a minimal extension of the SM in which an extra $U(1)$ dark sector is spontaneously broken by a new scalar field $\sigma$, see e.g.\ Refs.~\cite{Gonderinger:2009jp, Kadastik:2011aa, Elias-Miro:2012eoi, Chen:2012faa}. Well known examples are provided by the majoron~\cite{Chikashige:1980ui} and dark photon models~\cite{Galison:1983pa, Holdom:1985ag}, which can both related to FOPTs and detectable GWB.

In this {\it Letter}, we focus on the majoron model
\cite{Chikashige:1980qk,Gelmini:1980re,Schechter:1981cv,Akhmedov:1992hi}, as a benchmark for a wider class of FOPTs models. This amounts to a minimal extension of the SM symmetry group, which encodes an extra $U_{L}(1)$, and to the introduction of a new complex scalar singlet field $\sigma=\phi+\imath \, J$, with lepton charge $L:\sigma=-2$. Then $\sigma$ can be coupled to a $|\Delta L|=2$ Majorana operator $\nu_{L}\nu_{L}$ by considering a Lagrangian component\footnote{We set $\hbar = c = 1$ unless otherwise specified.} 
\begin{equation}
    \mathcal{L} \supset h_{L}\sigma\nu_{L}\nu_{L}+{\rm h.c.}\,,
\end{equation}
which does not violate the $U_{L}(1)$ symmetry. Extending the majoron model to include a type-I see-saw mechanism with a RH neutrino, an operator of the type
\begin{equation}
    \mathcal{L} \supset h_{R}\sigma \nu_{R}\nu_{R}+{\rm h.c.} \,,
\end{equation}
must be included in the Lagrangian.

The scalar sector of the model, including the SM Higgs field $H$, is subjected to the potential 
\begin{equation}
\label{Pot}
V_{0}(\sigma, H)=V_{0}'(\sigma, H)+V_{0}''(\sigma)+V_{0}'''(\sigma,H)\, , 
\end{equation}
where $V_{0}''$ and $V_{0}'''$ are higher order effective operators that can efficiently induce FOPTs in the dark scalar sector, while the potential for the Higgs and complex scalar singlet fields including the self-interactions is
\begin{eqnarray}
    V_{0}' &=& V_{0\sigma}+V_{0H}+V_{0\sigma H}\,,\label{Potzero}\\
    V_{0\sigma} &=& \lambda_{s}\Big(|\sigma|^{2}-\frac{v'^{2}}{2} \Big)^{2}\,, \label{sigma}\\
    V_{0H} &=& \lambda_{H}\Big(|H|^{2}-\frac{v^{2}}{2} \Big)^{2}\,, \label{HH}\\
    V_{0\sigma H} &=& \lambda_{\sigma H}\Big(|\sigma|^{2}-\frac{v'^{2}}{2} \Big)\Big(|H|^{2}-\frac{v^{2}}{2} \Big)\,.\label{VsH}
\end{eqnarray}
Assuming $\lambda_{\sigma H}<\!\!<\lambda_{\sigma,H}$ entails the suppression of the $\sigma-H$ mixing terms. We will assume that the fields $\sigma$ and $H$ are weakly coupled, which preserves $\sigma$ to be a dark scalar. 

The $V_{0\sigma}$ component of the potential entails the condensate's value $\langle \sigma \rangle=v'/\sqrt{2}$, which spontaneously breaks $U_{L}(1)$, generating a Majorana mass for neutrino mass term,
\begin{equation}
    \mathcal{L} \supset \mu_{L,R} \nu_{L,R} \nu_{L,R}+ {\rm h.c.}\,,    
\end{equation}
with $\mu_{L,R}=h_{L,R}\langle \sigma \rangle$. The pseudo Nambu-Goldstone boson related to the spontaneous symmetry breaking $J={\rm Im}\,\sigma$ is the majoron. 
The advantage of this mechanism is that it can be falsified by resorting to a multi-messenger strategy that also involves collider physics experiments   \cite{Das:2012ii, Deppisch:2013cya, Deppisch:2015qwa, CMS:2022qva}. The majoron scenario can also be tested through neutrino-less double beta decays ($0\nu\beta\beta$), thanks to the fact that the majoron emission should alter the distribution of the emitted electrons \cite{Blum:2018ljv, Cepedello:2018zvr}.

Considering only 6D L-preserving operators, the relevant potential terms expressed as higher order effective operators are
\begin{eqnarray}
    V_{0}'' &=& \frac{\kappa_{1}}{\Lambda^{2}}(\sigma^{*}\sigma)^{3}+{\rm h.c.}\,,\label{Vone}\\
    V_{0}''' &=& \frac{\kappa_{2}}{\Lambda^{2}}(HH)^{2}(\sigma^{*} \sigma)\!+\!\frac{\bar{\kappa}_{2}}{\Lambda^{2}}(HH)(\sigma^{*} \sigma)^{2}+ {\rm h.c.}\,.\label{Vd}
\end{eqnarray}
A further extension to the higher order terms of the potential can be envisaged by including terms which softly break $U_{L}(1)$~\cite{Addazi:2020zcj}. On top of that, the leading tree-level thermal corrections to the complex singlet mass to be considered is expressed by 
\begin{equation}
\label{deltaM}
\Delta M_{\sigma}(T)\simeq -6KT^{2}, \qquad  K=\frac{\kappa_{1}v'^{2}+\kappa_{2}v^{2}}{\Lambda^{2}}\,  .
\end{equation}
Thus, since the thermal contribution to the mass is negative, it can induces a FOPT from the $\sigma$-potential~\cite{Addazi:2020zcj,Addazi:2017nmg,Addazi:2017oge}. 

If $\kappa_{1}\simeq \kappa_{2}$, the second term in $K$ dominates over the first, with $v'<\!\!<v$, which is a necessary condition to derive a FOPT temperature compatible with GW observation by NANOGrav. This scenario can be eventually constrained through collider physics, from the Higgs decays into invisible channels. Nonetheless, the case $\kappa_{2}v^{2}<\!\!<\kappa_{1}v'^{2}$  also represents a viable possibility, as it eludes any direct constraint from colliders \cite{Das:2012ii, Deppisch:2013cya, Deppisch:2015qwa, CMS:2022qva}. We can then assume that the new scale of physics $\Lambda$ only relates to the majoron sector, being by assumption decoupled from the Higgs. The new physics scale $\Lambda$ can in principle be smaller than the electroweak scale.

Furthermore, in order to have a strong FOPT, $v'/\Lambda\simeq 0.1 \div 0.5$ can be safely considered without any violation of direct constraints and perturbative bounds. 
For this reason, in the analysis that follows, we are allowed to neglect the $\sigma-H$ interaction, hence providing a simplified explanation for the NANOGrav excess of stochastic spectrum. 

To proceed with the analysis of FOPTs, we construct the one-loop thermal corrected potential, adopting the same prescription as in Refs.~\cite{Quiros:1999jp}, namely
\begin{equation}
\label{Thermal}
V_{\rm eff}(T)=V_{0}+V_{\rm CW}^{(1)}+\Delta V(T)+V_{\rm C.T.}\, , 
\end{equation}
where $V_{0}$ is the tree-level potential in Eq.~\eqref{Pot}, $V_{\rm CW}$ is the 1-loop zero-temperature Coleman-Weinberg potential, $\Delta V(T)$ includes all the leading order thermal corrections, and the counter-term potential $V_{\rm C.T.\ }$ is written with the same prescription adopted in Ref.~\cite{Addazi:2023ftv}. In particular, the CW potential in the Landau gauge is expressed by  
\begin{equation}
\label{CW}
V_{\rm CW}^{(1)}=\sum_{i}(-1)^{F_{i}}n_{i}\frac{m_{i}^{4}}{64\pi^{2}}\Big[ \log\Big(\frac{m_{i}^{2}}{Q^{2}}\Big)-k_{i}\Big]\, , 
\end{equation}
where $n_{i}$ are the degrees of freedom of the system, $m_{i}\equiv m_{i}(\sigma, H)$ are field dependent particle masses --- the $i$-index runs over all the particles --- the exponents $F=0,1$ for bosons and fermions respectively, $Q$ denotes the renormalisation scale in the $\bar{MS}$-scheme, $k_{i}=1/2$ for transversely polarized gauge bosons and $k_{i}=3/2$ for longitudinally polarized gauge bosons, scalars and fermions. 
The one-loop thermal corrections can be expressed as 
\begin{equation}
\label{Thermal1}
\Delta V(T)=\frac{T^{4}}{2\pi^{2}}\Big\{\sum_{B}n_{B}J_{B}\Big[\frac{m_{B}^{2}}{T^{2}}\Big]-\sum_{F}n_{F}J_{F}\Big[\frac{m_{F}^{2}}{T^{2}}\Big] \Big\}\,,
\end{equation}
where $J_{B,F}$ are the boson and fermion thermal integrals, respectively. Eq.~\eqref{Thermal1} also contains 
the mass corrections in Eq.~\eqref{deltaM} as the leading order thermal contributions. 

The numerical analysis, performed by some of us in Refs.~\cite{Addazi:2017nmg,Addazi:2020zcj}, deployed the open source software {\it Cosmotransition}~\cite{Wainwright:2011kj}. This provides the numerical evaluation of the classical effective action $S_{3}\equiv S_{3}(\bar{\sigma}, \bar{H}, T)$, defined in terms of the full one-loop effective potential in Eq.~\eqref{Thermal}, by providing particular solutions $\bar{\sigma}$ and $\bar{H}$ that minimize the action~\cite{Coleman:1977py}.  

The thermal effective potential and the Euclidean action are involved in the definition of the $(\alpha, \beta)$ parameters, which are in turn related to FOPTs and GW spectra. Specifically, the latent energy parameter $\alpha$ is expressed by
\begin{equation}
\label{alpha}
\alpha=\frac{1}{\rho_{\gamma}}\Big[\Delta V-\frac{T}{4}\Big(\frac{\partial \Delta V}{\partial T}\Big)\Big]\, , 
\end{equation}
where $\Delta V=V_{f}(\sigma_{f}, H_{f}, T_{*})-V_{i}(\sigma_{i}, H_{i}, T_{*})$ is the difference between the potential in the final stable phase and the initial metastable phase, $T_{*}$ denoting a critical temperature;\footnote{The critical temperature $T_{*}$ for a FOPT to take place can be estimated either from the bubble nucleation temperature $T_n$, i.e.\ the temperature at which the rate of bubble nucleation per Hubble volume and time is approximately one, or more accurately from the percolation temperature $T_p$, at which the probability to have the false vacuum is about 0.7. Although using $T_p$ to calculate the strength leads to stronger signals for GWs and more visible results, here we conservatively estimate $T_{*}$ using $T_n$.} $\rho_{\gamma}=(\pi^{2}/30)g_{*}T_{N}^{4}$ is the radiation energy density during the epoch of bubble nucleation, $g_{*}$ denoting the number of relativistic degrees of freedom. The second characteristic FOPT parameter is $\beta$, which is the inverse of the typical average nucleation time scale
and is defined as:
\begin{equation}
\label{Defbeta}
\frac{\beta}{H_*}=T_{*}\frac{\partial}{\partial T}\Big(\frac{S_{3}}{T} \Big)\Big|_{T=T_{*}}\,.
\end{equation}

The three main contributions to GWs --- see e.g.\ Ref.~\cite{Caprini:2015zlo} --- from FOPTs are: i) bubble-bubble deep inelastic scatterings; ii) shock sound waves (sw); iii) turbulence of the primordial plasma. These GWs are sourced in a fast transient around the bubble nucleation time, and due to the Universe expansion undergo redshift to observers at present time. 

The predominance of one contribution upon the others depends crucially on the nucleation dynamics. This is in turn characterised by the bubble speed profile $v_{B}$, as a function of the parameter $\alpha$~\cite{Caprini:2015zlo}. In particular, for the runaway bubbles, characterized by $v_{B}\approx 1$, the collision contribution is dominant; while for the non-runaway bubbles, the sound and turbulence sources are the leading order contributions toe the GW spectra.

For the scientific case we are representing in this {\it Letter}, it is essential to remark that any $\Omega_{\rm GW}(f)$ contribution determines different frequency-dependent power-spectra profiles. This ensures the falsifiability of the proposed models through the comparison of the spectra predicted by the PTA analyses. The statistical analyses of the best fit then enable to disentangle, as a inverse problem, the nature of the FOPT, clarifying the scenario, either runaway or not runaway, favoured by the recent PTA results. \\

\textbf{Results.} 
The spectral shape of the GWB is phenomenologically described in terms of an approximated broken power law function of the form
\begin{equation}
\mathcal{S}=\frac{1}{\mathcal{N}} \frac{(a+b)^c}{b x^{-a/c} + a x^{b/c}}\,,
\end{equation}
in which 
\begin{equation}
\mathcal{N}= \left(\frac{b}{c} \right)^{a/n} \, \left(\frac{n c}{b} \right)^{c} \, \frac{\Gamma(a/n) \Gamma(b/n)}{n \Gamma(c)} 
\,,
\end{equation}
with $n=(a+b)/c$, and the spectral shape parameters $a$, $b$ and $c$ have been estimated from the data, selecting their values in prior ranges that account for the typical uncertainties of numerical simulations and the possible dependence on specific models --- see e.g.\ Refs.~\cite{Hindmarsh:2015qta, Ellis:2020awk}. These values are finally deployed to determine the spectral index, as we specify below.   

The GWB amplitude is modeled in terms of a power-law spectrum with the characteristic strain
\begin{equation}
	\label{eq:strain}
	h_c(f) = A_* \left(\frac{f}{f_{\rm yr}}\right)^{\frac{3-\gamma}{2}}\,,
\end{equation}
where $f_{\rm yr} = 1{\rm \, yr^{-1}}$ is a reference frequency, $A_*$ is the amplitude at $f_{\rm yr}$, and the parameter $\gamma$ is related to the spectral tilt. The fractional energy density in GWs associated with the strain is~\cite{Ellis:2020ena}
\begin{equation}
	\label{eq:strain1}
	\Omega_{\rm GW}(f) = \frac{2 \pi^{2}}{3 H_0^2}f^{2}h_c^2(f) \equiv \Omega_{\rm GW}^{\rm yr} \left(\frac{f}{f_{\rm yr}}\right)^{5-\gamma}\,,
\end{equation}
where $\Omega_{\rm GW}^{\rm yr} = 2 \pi^{2}f_{\rm yr}^{2}/(3 H_0^2)$. The constraint derived on the $(A_*, \gamma)$ space of parameters by the NANOGrav-$15\,$yr collaboration may be then interpreted as a GWB signal of amplitude $A_* = 6.4_{-2.7}^{+4.2}\times 10^{-15}$ at 90\% credibility and with spectral index $\gamma = 3.2_{-0.6}^{+0.6}$~\cite{NANOGrav:2023gor}. PPTA consistently reports the amplitude $A_* = 3.1_{-0.9}^{+1.3}\times 10^{-15}$ at 68\% credibility for the spectral index $\gamma = 3.90\pm 0.40$~\cite{Reardon:2023gzh}. The analysis by the EPTA collaboration yields the amplitude $\log_{10}A_* = -14.54_{-0.41}^{+0.28}$ at 90\% credibility, with the spectral index $\gamma = 4.19_{-0.63}^{+0.73}$ when using the full set of data covering 24.7 years~\cite{Antoniadis:2023ott}. In the following, we account for the results by the NANOGrav-$15\,$yr collaboration as a benchmark of the recent PTA data release.

We assess the cosmological scenario presented against these experimental results, considering the GW signal from phase transitions. We obtain the spectral tilt and the amplitude by inverting the relations in Eqs.~\eqref{eq:strain}-\eqref{eq:strain1} as in Ref.~\cite{Ellis:2020ena}, namely 
\begin{eqnarray}
	\gamma &=& 5 - \frac{\mathrm{d}\ln\Omega_{\rm GW}(t_0, f)}{\mathrm{d}\ln f}\bigg|_{f=f_*}\,,\\
	A_* &=& \sqrt{\frac{3H_0^2}{2\pi^2}\,\frac{\Omega_{\rm GW}(t_0, f_*)}{f_{\rm yr}^2}\,\left(\frac{f_{\rm yr}}{f_*}\right)^{5-\gamma}}\,,
\end{eqnarray}
where the quantities are computed at the reference frequency $f_* = f_{\rm yr}$.

For frequencies in the nHz region, the dominant contribution is set by shock sound waves, whose dynamics has been simulated in Refs.~\cite{Hindmarsh:2015qta, Ellis:2020awk}. A fit to the simulation in Ref.~\cite{Hindmarsh:2015qta} leads to the spectral index
\begin{equation}
    \gamma = 2 + 21 f^2/(3 f^2 + 4 f_{\rm sw}^2)\,,
\end{equation}
where the peak frequency is set by the average bubble separation $f_{\rm sw} \approx \beta/v_B$. Once redshift is taken into account, the value of the characteristic frequency only depends upon the combination $\beta T_*/H_*$, so that the 15\,yr result by the NANOGrav collaboration can be recast in terms of the parameters describing the FOPT as
\begin{equation}
    2.4 \lesssim \frac{1}{v_B}\frac{\beta}{H_*} \frac{T_*}{\rm 100\,MeV} \lesssim 4.3\,.
\end{equation}

\begin{figure}
	\includegraphics[width = \linewidth]{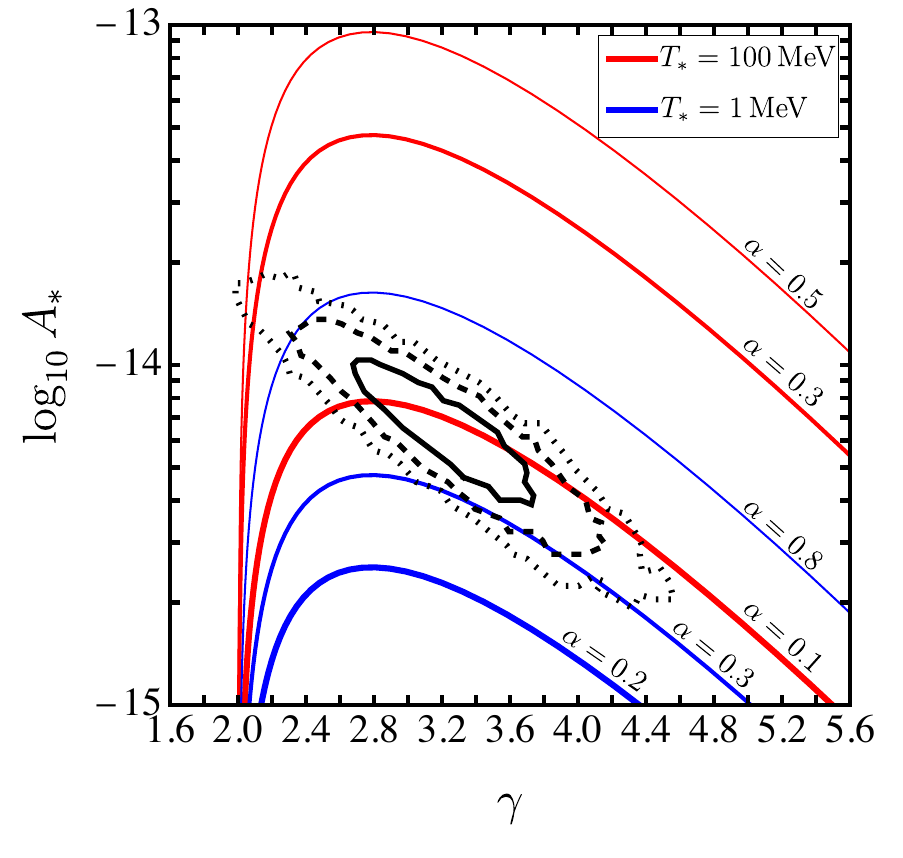} 
	\caption{The strain amplitude $A_*$ (vertical axis) and the index $\gamma$ (horizontal axis) predicted by different models of GWs from bubble collision for a phase transition occurring at $T_* = 100\,$MeV (red lines) and $T_* = 1\,$MeV (blue lines). Different line thicknesses correspond to different choices for the value of $\alpha$ --- see the figure label for details. We also display the fit recently obtained by the NANOGrav collaboration~\cite{NANOGrav:2023gor} after analyzing the 15 years data set, with the different shadings marking the 1-, 2- and 3-$\sigma$ confidence regions.}
	\label{fig:bubble}
\end{figure}

In Fig.~\ref{fig:bubble}, we compare the parametric region $(A_{*},\gamma)$ from NANOGrav results with several FOPTs benchmarks, picking different values of $\alpha$ and $T_{*}$. As displayed, there exist several FOPTs with $T_{*}\simeq 1\div 100\,{\rm MeV}$ and $\alpha\simeq 0.1\div 0.3$ that are compatible with the NANOGrav excess of a stochastic spectrum. The degeneracy in the $(T_*, \alpha)$ parameter space has also been obtained in the recent Monte Carlo analysis by the NANOGrav collaboration~\cite{NANOGrav:2023hvm} and allows to accommodate the results by the PPTA and EPTA collaborations which report a different best fit within 2$\sigma$.

\textbf{Conclusions.} In this {\it Letter}, we have analyzed data recently released by pulsar timing array (PTA) collaborations within the working assumption of a first order phase transition (FOPT) being originated from a $U(1)$ dark sector. We have shown that a FOPT with a critical temperature around $1\div 100\, {\rm MeV}$ can provide a natural explanation for the excess of stochastic spectrum recently reported using the PTA methods. Differently than other possibilities, this model does not appear to be tailored {\it ad hoc} in order to fit the PTA results. In fact, while in Fig.~\ref{fig:bubble} we have focused on the results by the NANOGrav consortium which is here used as a benchmark, we remark that the results by the EPTA, PPTA and CPTA collaborations can also be accommodated within the model presented.

The nucleation temperature that fits the data lies close to the mass scale that is relevant for warm dark matter (WDM) models such as the majoron scenario. Within a multi-messenger perspective, radio-astronomy can provide a powerful probe for WDM models. It is anyway important to point out that also other models with spontaneously broken extra $U(1)$, including the dark photon, can in principle provide FOPTs~\cite{Addazi:2017gpt} compatible with the data released by NANOGrav. An exhaustive analysis of all the viable DM models from SM group extensions with FOPTs is beyond the purposes of this {\it Letter}.

\vspace{0.5cm}

\begin{acknowledgments}
The work of A.A.\ is supported by the Talent Scientific Research Program of College of Physics, Sichuan University, Grant No.\ 1082204112427.
Y.C.\ acknowledges the National Key R\&D Program of China (2021YFC2203100), CAS Young Interdisciplinary Innovation Team (JCTD-2022-20), NSFC (12261131497), 111 Project for ``Observational and Theoretical Research on Dark Matter and Dark Energy'' (B23042), Fundamental Research Funds for Central Universities, CSC Innovation Talent Funds, USTC Fellowship for International Cooperation, USTC Research Funds of the Double First-Class Initiative, CAS project for young scientists in basic research (YSBR-006).
A.M.\ acknowledges support by the NSFC, through the grant No.\ 11875113, the Shanghai Municipality, through the grant No.\ KBH1512299, and by Fudan University, through the grant No.\ JJH1512105.
This publication is based upon work from the COST Actions ``COSMIC WISPers'' (CA21106) and ``Addressing observational tensions in cosmology with systematics and fundamental physics (CosmoVerse)'' (CA21136), both supported by COST (European Cooperation in Science and Technology).
\end{acknowledgments}

\bibliographystyle{apsrev4-1}
\bibliography{refs.bib}

\end{document}